\documentclass[useAMS,usenatbib]{mn2e}

 \usepackage{times}
 \usepackage{graphicx}

\def\pcm3{{\rm\thinspace cm$^{-3}$}}

\def\contcaption{\@conttrue\SFB@caption\@captype}

\title[Proper motions of Upper Sco T--type candidates]{
Proper motions of USco T--type candidates
\thanks{Based on observations collected with the ESO New Technology Telescope
under programme number 089-C.0854(A).}
\thanks{Based on observations made with the Gran Telescopio Canarias (GTC),
installed in the Spanish Observatorio del Roque de los Muchachos of the
Instituto de Astrof'sica de Canarias, in the island of La Palma.}}
\author[N. Lodieu]{N. Lodieu$^{1,2}$\thanks{E-mail: nlodieu@iac.es},
V.\ D.\ Ivanov $^{3}$, \& P.\ D.\ Dobbie$^{4}$  \\
$^{1}$ Instituto de Astrof\'isica de Canarias (IAC), C/ V\'ia L\'actea s/n, 
E-38200 La Laguna, Tenerife, Spain \\
$^{2}$ Departamento de Astrof\'isica, Universidad de La Laguna (ULL),
E-38205 La Laguna, Tenerife, Spain \\
$^{3}$ European Southern Observatory, Santiago de Chile, Chile \\
$^{4}$ School of Mathematics \& Physics, University of Tasmania, Hobart, TAS, 7001, Australia
}
\begin{document}

\date{Accepted \today. Received \today; in original form \today}

\pagerange{\pageref{firstpage}--\pageref{lastpage}} \pubyear{2005}

\maketitle

\label{firstpage}

\begin{abstract}
We present new $z$- and $H$-band photometry and proper motion measurements 
for the five candidate very-low-mass T--type objects we recently proposed 
to be members of the nearest OB association to the Sun, Upper Scorpius. 
These new data fail to corroborate our prior conclusions regarding their 
spectral types and affiliation with the Upper Scorpius population. 
We conclude that we may be in presence of a turnover in the mass
function of Upper Sco taking place below 10--4 Jupiter masses,
depending on the age assigned to Upper Sco and the models used.
\end{abstract}


\section{Introduction}
\label{USco_dT_PM:intro}

The quest for young objects of spectral-type T remains an area of 
substantial interest as a way to address a fundamental question in 
our understanding of star formation: what is the lowest mass that this 
process can form ? The earliest theoretical predictions by \citet{kumar69},
\citet{low76}, and \citet{rees76} suggested masses as low as $\sim$10 
Jupiter (M$_{\rm Jup}$) but contemporary calculations reveal that in 
the presence of magnetic fields this limit could be much lower 
\citep{boss01,stamatellos08}.

Naturally, the searches for these objects have concentrated on the 
nearest young clusters and star-forming regions and these have led to 
the identification of several candidate infantile T--type objects. 
Crucially, none of these has been unambiguously confirmed astrometrically 
{\it{and}} spectroscopically. For example, \citet{bihain10a} have 
detected a further candidate T--type member of the $\sigma$ Ori cluster, 
adding to the previously known candidate mid-T, S\,Ori\,70 
\citep{zapatero02b,zapatero08a,burgasser04b,scholz08a,luhman08c,zapatero08a}. 
However, proper motion measurements of both objects cast doubt on 
their association with this population \citep{penya11a}. More recently, 
\citet{penya12a} have identified another candidate T--type in this same 
region using photometry from the VISTA \citep[Visible and Infrared 
Survey Telescope for Astronomy;][]{emerson04} Orion survey.

Additionally, \citet{marsh10} have claimed the discovery of a T2 member 
of $\rho$ Ophiuchus but this has since been refuted by \citet{alves10}. 
Independently, \citet{geers11} has proposed several candidates as 
substellar members of this population through infrared spectroscopy, 
including one with a mass close to the deuterium burning limit. Another 
wide-field methane imaging survey of $\rho$ Ophiuchus revealed 22 T--type 
dwarf candidate members down to 1--2 Jupiter \citep{haisch10}. 
\citet{burgess09} identified a mid-T--type candidate from a deep 
methane survey of $\sim$0.11 square degrees in IC\,348 but neither 
spectroscopy nor astrometry is yet available to confirm membership. 
\citet{spezzi12b} reported two potential T--type candidates in the
core of the Serpens cloud although their nature remains uncertain
with the sets of data available to the authors.
Similarly, none of the faint Pleiades L/T dwarf candidates announced 
by \citet{casewell07} have been confirmed spectroscopically as members 
\citep{casewell11}. It is worth noting here that there are two 
spectroscopically and astrometrically confirmed T dwarf members of 
the Hyades cluster \citep{bouvier08a}. However, these have significantly 
larger masses ($\sim$50 Jupiter masses) than the young T--types due 
to their substantially greater ages, $\tau$$\sim$600\,Myr. 

Upper Scorpius (hereafter USco) is part of the Scorpius Centaurus OB 
association: it is located at 145 pc from the Sun \citep{deBruijne97} 
and its age is estimated to 5$\pm$2 Myr from isochrone fitting and 
dynamical studies \citep{preibisch99} although a more recent study by 
\citet{pecaut12} suggests 11$\pm$2 Myr \citep[see also][]{song12}. The 
association has been targeted in X~rays 
\citep{walter94,kunkel99,preibisch98}, astrometrically with Hipparcos 
\citep{deBruijne97,deZeeuw99}, and more recently at optical 
\citep{preibisch01,preibisch02,ardila00,martin04,slesnick06} and 
near--infrared \citep{lodieu06,lodieu07a,dawson11,lodieu11c,dawson12} 
wavelengths. Tens of brown dwarfs have now been confirmed 
spectroscopically as members of USco
\citep{martin04,slesnick06,lodieu06,slesnick08,lodieu08a,martin10a,lodieu11a}
 and the mass function of this population determined robustly, deep 
into the substellar regime \citep{preibisch02a,slesnick08,lodieu11a}.

In a recent paper in our extensive series of publications relating to 
USco \citep{lodieu11a} we identified five T--type candidate members 
with deep infrared photometry from the UK Infrared telescope wide-field 
camera \citep[UKIRT/WFCAM;][]{casali07}. In the current work, we report 
proper motion measurements for these objects obtained from early deep 
WFCAM $J$-band observations and new $H$-band imaging that are separated 
in time by four years. In Section \ref{USco_dT_PM:IRphot} and 
Section \ref{USco_dT_PM:OPTphot} we describe the new $H$-band 
observations carried out with the Son of Isaac (SofI) instrument 
installed on the European Southern Observatory (ESO) New Technology 
Telescope (NTT) in La Silla Observatory (Chile) and additional $z$-band 
imaging conducted with the Optical System for Imaging and low Resolution 
Integrated Spectroscopy (OSIRIS) installed on the Gran Telescopio de 
Canarias (GTC) in La Palma Observatory (Canary Islands). In 
Section \ref{USco_dT_PM:membership} we use the new photometry and 
astrometry to examine the membership status of the five candidate 
T--type members. In Section \ref{USco_dT_PM:discussion} we place our 
new results into context and speculate about our (positive/negative) 
results.

\section{Near-infrared photometry}
\label{USco_dT_PM:IRphot}

\subsection{$H$-band imaging}
\label{USco_dT_PM:IRphot_Hband}

We performed near-infrared imaging of the five T--type candidates in USco
listed in Table 4 of \citet{lodieu11a} in the $H$ filter with 
SofI on the 3.5-m NTT \citep{moorwood98}. All five sources were observed 
on 9 May 2012\@.

SofI is equipped with a Hawaii HgCdTe 1024$\times$1024 array with squared 
18.5 micron pixels and has both imaging and spectroscopic capabilities. 
The pixel scale is 0.292 arcsec in the Large Field configuration, 
providing coverage of a 4.9$\times$4.9 arcmin field-of-view. We employed 
a random dithering pattern within a 40 arcsec box using six on-source 
individual integrations of 20 sec, repeating this 10 or 20 times for the 
two brightest and three faintest T--type candidates. This yielded total 
on-source exposures of 20 or 40 min for the bright and faint objects, 
respectively. At the time of the observations, the night was clear and 
the seeing was around 0.7--0.9 arcsec, allowing us to go deep enough 
to detect the T--type candidates with signal-to-noise ratios between 
8 and 15\@. Dome flats and darks with the same on-source integrations 
were taken during the afternoon prior to our observing night. As our 
fields are within the footprint of the UKIRT Infrared Deep Sky Survey 
Galactic Clusters Survey \citep[UKIDSS GCS;][]{lawrence07}, no photometric standard stars were observed. 

\subsection{Data reduction and astrometry}
\label{USco_dT_PM:IRphot_DR}

The data were reduced with the ESO EXOREX SofI pipeline recipes. 
These perform an automatic reduction of the target frames 
within an observing block, including flat field correction, sky 
subtraction, and cross-talk removal. The next step of the data analysis 
included photometric and astrometric calibrations using the WFCAM images 
as reference.

To astrometrically calibrate the SofI images we proceeded as follows:
for a first guess we used the astrometry.net package\footnote{More details at 
astrometry.net} which requires the centre of image given by the (RA,dec) 
coordinates in the header, the pixel scale (0.292 arcsec/pixel), and a 
radius for the search (set to 12 arcmin, more than twice the field-of-view 
of the SofI images). The astrometric solution was satisfactory comparing 
with 2MASS and the deep WFCAM images obtained as first epoch. However, it 
was not good enough for our purposes, i.e.\ to measure proper motions 
between the two epochs.

The second step made use of the GAIA software\footnote{GAIA is a 
derivative of the Skycat catalogue and image display tool, developed 
as part of the VLT project at ESO. Skycat and GAIA are free software 
under the terms of the GNU copyright.}
which itself uses SExtractor \citep{bertin96}. We ran the detection 
algorithm to extract all sources (pixel and world coordinates systems) 
in the SofI images. Then, we cross-correlated this SExtractor 
catalogue against the deep WFCAM dataset and kept only the SofI (x,y) and
WFCAM (RA,dec) coordinates in an output file for sources with $J$-band
magnitudes in the 19--20 range. Next we used the IRAF task {\tt{ccmap}} 
interactively with a polynomial of order four. Using the faintest stars 
from the WFCAM images allowed us to exploit more than 100--170
point sources with a small intrinsic motion on the sky (i.e.\ about
21--25\% of all stars in the each SofI field), avoiding bright
members of the association. We eliminated points whose astrometry was off 
by more than 5$\sigma$, yielding an rms of 44.8--51.1 mas and 
33.2--45.1 mas in right ascension and declination, respectively
(corresponding to about 1/6 of the SofI pixel scale or 11--13 mas/yr). 
The new image was saved and SExtractor ran again with a detection 
threshold of 3$\sigma$ and an aperture twice the size of the 
full-width-half-maximum ($\sim$6 pixels or 2 arcsec) to detect all 
sources in the SofI field-of-view, including the targets.

\subsection{Photometric calibration}
\label{USco_dT_PM:IRphot_Photometry}

We could not use point sources within the 2MASS database to calibrate 
photometrically the SofI frames because most of these were saturated in 
our images. Instead, we cross-matched all objects detected by SExtractor 
(see Section \ref{USco_dT_PM:IRphot_DR}) with the ninth data release
of the UKIDSS GCS and retrieved all point sources detected in $H$
with photometric error bars less than 0.1 mag for each individual field. 
The total numbers of matched sources within a matching radius of two 
arcsec was typically 200--240\@. We find a median offset of 
$-$0.799$\pm$0.095 mag between the UKIDSS system 
\citep[Vega system;][]{hewett06} and the SofI photometry when we adopt 
the default zero point of 25 mag in the SExtractor parameter field. 
We list the offsets for each of the five fields in 
Table \ref{tab_USco_dT_PM:ZP_Hband}. The final photometric uncertainties 
on the offsets corresponds to the root mean square of the dispersion 
between offsets and the individual errors. 
Table \ref{tab_USco_dT_PM:photometry_dT} lists the $H$ magnitudes and 
their errors of our five USco targets, computed using the offsets from 
each individual frame.

\begin{table}
 \centering
 \caption[]{Offsets between the NTT/SofI and UKIRT/WFCAM $H$-band 
photometry using $>$100 point sources in each individual SofI field.
The last row indicates the mean (Avg) value of the offset, taking into 
account the dispersion and errors on the individual offsets.
}
 \begin{tabular}{@{\hspace{0mm}}c c c c c@{\hspace{0mm}}}
 \hline
 \hline
Field & R.A.    &     Dec       &   Offset ($H$)     & \# stars \cr
 \hline
      & hh:mm:ss.ss & ${^\circ}$:$'$:$''$ & mag &   \cr
 \hline
1 & 16:08:35.98 & $-$22:29:11.1 & $-$0.796$\pm$0.082 & 210 \cr
2 & 16:08:45.73 & $-$22:29:53.5 & $-$0.784$\pm$0.064 & 226 \cr
3 & 16:08:47.80 & $-$22:29:04.5 & $-$0.846$\pm$0.070 & 243 \cr
4 & 16:09:55.91 & $-$22:33:45.7 & $-$0.807$\pm$0.070 & 203 \cr
5 & 16:10:04.76 & $-$22:32:30.6 & $-$0.761$\pm$0.080 & 197 \cr
 \hline
Avg &             &               & $-$0.799$\pm$0.095 &     \cr
  \hline
 \label{tab_USco_dT_PM:ZP_Hband}
 \end{tabular}
\end{table}

\begin{table*}
 \centering
 \caption[]{Photometry for the USco T--type candidates: the
 $Y,J$ and methane photometry is from \citet{lodieu11a} to which we added
 the new $H$-band photometry from NTT/SofI and $z$-band magnitudes
from GTC/OSIRIS. The $z-J$ and $J-H$ colours are given as well. The 
resulting proper motions measured between the first
and second epoch images at near-infrared wavelengths are quoted in mas/yr.}
 \begin{tabular}{@{\hspace{0mm}}c @{\hspace{1.5mm}}c @{\hspace{1.5mm}}c @{\hspace{1.5mm}}c @{\hspace{1.5mm}}c @{\hspace{1.5mm}}c @{\hspace{1.5mm}}c @{\hspace{1.5mm}}c @{\hspace{1.5mm}}c @{\hspace{1.5mm}}c @{\hspace{1.5mm}}c @{\hspace{1.5mm}}c @{\hspace{1.5mm}}c@{\hspace{0mm}}}
 \hline
 \hline
R.A.        &     Dec      &  $Y$  &  $J$   & CH$_{\rm 4off}$ & CH$_{\rm 4on}$ & $H$ & $z$ & $J-H$ & $z-J$ & $\mu_{\alpha}\cos{\delta}$ & $\mu{\delta}$ & $\mu$ \cr
 \hline
hh:mm:ss.ss & ${^\circ}$:$'$:$''$ & mag & mag & mag & mag & mag & mag & mag & mag & mas/yr & mas/yr & mas/yr \cr
 \hline
16:08:35.98 & $-$22:29:11.1 & 21.87$\pm$0.14 & 20.96$\pm$0.11 & 20.47$\pm$0.20 & 21.11$\pm$0.18 &  20.41$\pm$0.15 & 22.59$\pm$0.04 & 0.55$\pm$0.18 & 1.63$\pm$0.12 & $-$3.6 & $+$12.8  & 13.3 \cr
16:08:45.73 & $-$22:29:53.5 & 21.47$\pm$0.10 & 20.72$\pm$0.09 & 20.29$\pm$0.17 & 21.15$\pm$0.13 &  20.20$\pm$0.12 & 22.37$\pm$0.07 & 0.52$\pm$0.15 & 1.65$\pm$0.11 & $-$18.4 & $+$6.9  & 19.7 \cr
16:08:47.80 & $-$22:29:04.5 & 21.57$\pm$0.11 & 20.72$\pm$0.09 & 20.59$\pm$0.24 & 21.45$\pm$0.17 &  20.28$\pm$0.13 & 22.48$\pm$0.08 & 0.44$\pm$0.16 & 1.76$\pm$0.12 & $+$12.9 & $-$5.8  & 14.1  \cr
16:09:55.91 & $-$22:33:45.7 & 21.10$\pm$0.08 & 20.19$\pm$0.06 & 19.81$\pm$0.14 & 20.40$\pm$0.08 &  19.87$\pm$0.16 & 22.02$\pm$0.07 & 0.32$\pm$0.17 & 1.83$\pm$0.09 & $+$21.0 & $-$12.7 & 24.5  \cr
16:10:04.76 & $-$22:32:30.6 & 20.46$\pm$0.05 & 19.69$\pm$0.04 & 19.26$\pm$0.11 & 19.75$\pm$0.04 &  19.16$\pm$0.12 & 21.50$\pm$0.09 & 0.52$\pm$0.12 & 1.81$\pm$0.10 & $+$35.4 & $+$26.3 & 44.1  \cr
  \hline
 \label{tab_USco_dT_PM:photometry_dT}
 \end{tabular}
\end{table*}

\begin{figure*}
  \centering
  \includegraphics[width=\linewidth, angle=0]{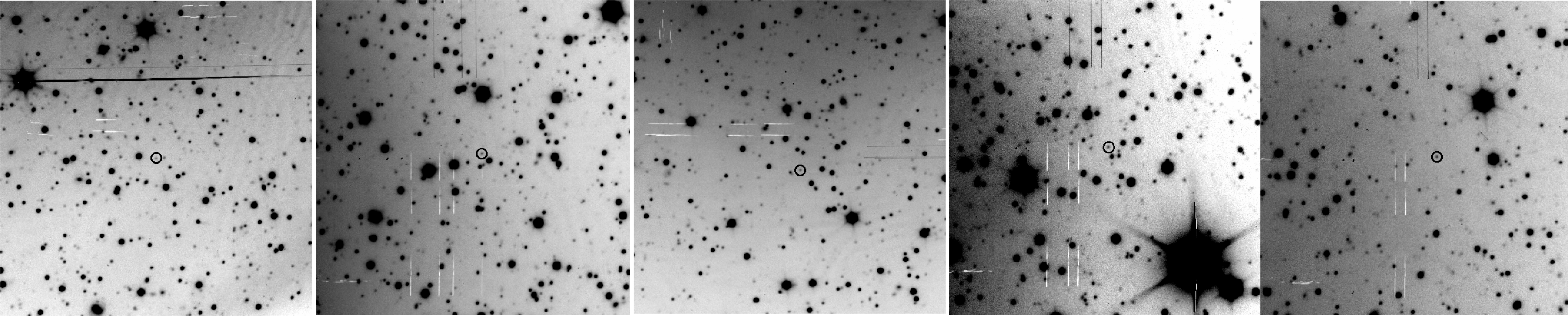}
  \caption{GTC/OSIRIS $z$-band images for the five candidates (circled 
and centered). North is up and East is left. Images are 1$'$$\times$1$'$.}
  \label{fig_dT_PM:FC_zband}
\end{figure*}

\section{Optical photometry}
\label{USco_dT_PM:OPTphot}

\subsection{$z$-band imaging}
\label{USco_dT_PM:OPTphot_zband}

OSIRIS is the Optical System for Imaging and low Resolution Integrated
Spectroscopy instrument \citep{cepa00} on the 10.4-m GTC operating at the 
Observatory del Roque de Los Muchachos (La Palma, Canary Islands).
The OSIRIS instrument is equipped with two 2048$\times$4096 Marconi 
CCD42-82 with a 8 arcsec gap between them and operates at optical 
wavelengths, from 365 to 1000 nm. The unvignetted instrument 
field-of-view is about 7$\times$7 arcmin with a pixel scale of
0.125 arcsec. We used the standard 2$\times$2 binning mode.

We imaged the five T--type candidates in USco with the Sloan $z$ filter 
available on OSIRIS during May 2012\@. Bias and skyflats were observed on 
26 May (evening) and 28$+$30 May (morning). On 27 May 2012, we obtained 
three series of 10 frames with 60 sec on-source integrations covering 
the targets 16084780$-$2229045, 16084573$-$2229535 and 
16083598$-$2229111. On 29 May 2012, we obtained three sets of 10 images 
with 20 sec on-source integrations for 16095591$-$2233457 and 
16100476$-$2232306 as well as nine images of 39 sec for 
16100476$-$2232306\@. We also repeated the observations of 
16084780$-$2229045 and 16084573$-$2229535 obtaining four series of 
10 images exposed 60 sec.

All observations were conducted under average seeing of 1.1--1.3 arcsec, 
photometric or clear conditions, and airmass between 1.6 and 1.8\@. The 
sky was relatively dark during the observations made on 27 May because 
the moon was set whereas the 64\%-full moon was below 20${^\circ}$ on 
29 May 2012\@.

\subsection{Data reduction and astrometry}
\label{USco_dT_PM:OPTphot_DR}

We reduced the OSIRIS Sloan $z$-band images in a standard manner under 
the IRAF environment \footnote{IRAF is distributed 
by the National Optical Astronomy Observatories, which are operated by 
the Association of Universities for Research in Astronomy, Inc., under 
cooperative agreement with the National Science Foundation} 
\citep{tody86,tody93}. First, we subtracted the mean bias and divided 
by the normalised averaged master skyflat to each individual science 
frame. Then, we combined each set of 10 images taken without dithering
and finally combined those sets applying the offsets to create a master
science frame. We note that our targets were located on CCD \#2, 
thus, we only treated data from that chip throughout the reduction process.

We calibrated astrometrically the final combined science frames using 
IRAF and {\tt{ds9}} \citep{joye03}. First, we saved in a file a list of 
point sources from the 2MASS catalogue \citep{cutri03,skrutskie06} 
spread over the full OSIRIS field-of-view. Second, we ran the 
{\tt{daofind}} task with the adequate detection and threshold parameters 
to identify (roughly) the same point sources to cross-match them in a 
subsequent step with the {\tt{ccxymatch}} routine. The latter task 
required a reference star with pixel (x,y) and world coordinate system 
(ra,dec) coordinates to efficiently cross-match the 2MASS (x,y) and 
(ra,dec) catalogues. We typically found 80--100 stars in the 
field-of-view of CCD \#2 running {\tt{ccmap}} with a polynomial of
order four, resulting in an astrometric calibration better than 
0.1--0.15 arcsec. 
The final reduced $z$-band images of the five candidates are shown in 
Fig.\ \ref{fig_dT_PM:FC_zband}.

\subsection{Photometric calibration}
\label{USco_dT_PM:OPTphot_Photometry}

The GTC calibration plan provided us with only one observation of a 
photometric standard star (G\,163-50) taken on the night of 27 May 2012 
with a single on-source integration of 0.8 sec at an airmass of 1.253\@. 
This DA3.2 white dwarf \citep{holberg12} is a Sloan photometric standard 
\citep{adelman_mccarthy11} and has a $z$-band magnitude of 13.809\@. 
We measured the instrumental magnitude using aperture photometry and applied 
a curve-of-growth analysis to allow for all the flux from the standard star.

We obtained a photometric zero point of 28.028$\pm$0.020 mag which is 
consistent within the error bars with both the values from the GTC OSIRIS 
daily monitoring of the zero points\footnote{www.gtc.iac.es/en/media/osiris/zeropoints.html} and 
our own previous measurement (28.038$\pm$0.059) from data taken in 
Semester 12B (Lodieu et al.\ 2012, submitted to A\&A).
For the night of 29 May 2012, we use the average value from Semester 12B
quoted above although data from the Carlsberg Meridian Telescope 
(http://www.ast.cam.ac.uk/ioa/research/cmt/data/camcext.12) suggests 
this night was similar in transparency to the first.

We performed aperture and point-spread function (PSF) photometry with
{\tt{daophot}} under IRAF because of the fairly crowded nature of this 
region (Fig.\ \ref{fig_dT_PM:FC_zband}) and the
faintness of our targets. We choose an aperture equal to 3$\times$ the
full-width-at-half-maximum and checked that our targets were all well
subtracted without residuals by our PSF analysis. We corrected the 
instrumental magnitude for the $z$-band zero point and the airmass.
We did not take into account possible effects due to colour terms.
We list in Table \ref{tab_USco_dT_PM:photometry_dT} the final magnitudes 
of the five T--type candidates in USco. We note that we quote the mean 
value of the magnitudes when two measurements were available (case of 
16084573$-$2229535, 16084780$-$2229045, and 16100476$-$2232306), the 
uncertainty being the dispersion between both values to which we added 
in quadrature.

\begin{figure}
  \includegraphics[width=\linewidth, angle=0]{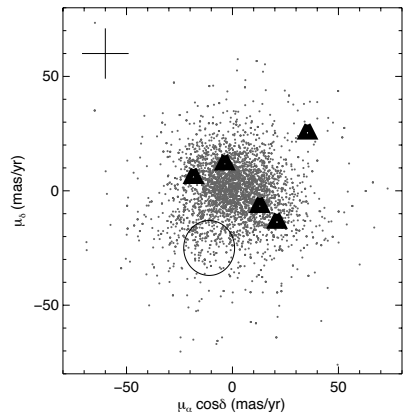}
  \caption{Proper motion vector point diagrams for the five T--type
candidates in USco marked with thick black triangles. 
The large circle has a radius of 12 mas/yr and is centered on the
USco mean proper motion. The small grey dots represent all point 
sources common to the deep WFCAM survey and the NTT fields.}
  \label{fig_dT_PM:VPdiagram}
\end{figure}

\section{Re-examining membership to USco}
\label{USco_dT_PM:membership}

\subsection{New astrometric tests}
\label{USco_dT_PM:membership_PM}
To measure the relative proper motions for all common point sources,
we cross-matched the catalogues from the five NTT pointings with the 
full catalogue of the deep WFCAM survey \citep{lodieu11c} with a 
matching radius of two arcsec.  
We found about 3200 sources to compare with the proper motions measured
for the five T--type candidates. We list the proper motion in the right 
ascension and declination as well as the total proper motion in
Table \ref{tab_USco_dT_PM:photometry_dT}.  We show their positions in 
proper motion reduced vector point diagrams in 
Fig.\ \ref{fig_dT_PM:VPdiagram} where our T--type candidates are 
highlighted with thick black triangles. All five candidates lie 
at least 2.5$\sigma$ from the mean absolute proper motion of the 
association estimated as ($-$11,$-$25) mas/yr by Hipparcos 
\citep{deBruijne97,deZeeuw99}, arguing against their membership to the 
association.

We compiled a list of known spectroscopic members of USco from 
\citet{ardila00}, \citet{martin04}, \citet{slesnick06}, \citet{lodieu06}, 
\citet{slesnick08}, \citet{dawson11}, \citet{lodieu11a}, and 
\citet{dawson12} to cross-match with the catalogue of point sources 
common to the NTT fields-of-view and the deep WFCAM survey 
\citep{lodieu11c}. 
Unfortunately, none of these known spectroscopic members lies within the 
NTT fields-of-view. This is not surprising considering that the total 
area covered by the five NTT pointings is of the order of 0.03
square degrees. \citet{lodieu06} and \citet{lodieu07a} found between 0.1
and 0.5 member candidates in 0.03 square degrees down to the depth of
the UKIDSS GCS, depending on the location in the association. 

\subsection{New photometric tests}
\label{USco_dT_PM:membership_phot}

Using our new photometry we have derived the $J-H$ colours of the five 
USco candidates so that we can further probe their nature 
(Table \ref{tab_USco_dT_PM:photometry_dT}). We find that these lie in 
the range 0.32--0.55 with an upper limit on the photometric errors of 
0.18 mag, implying that these sources could be either T2--T4 dwarfs or 
M dwarfs due to the degeneracy in the near-infrared colours 
\citep{hawley02,west05,hewett06,pinfield08}. 

\begin{figure}
  \centering
  \includegraphics[width=\linewidth, angle=0]{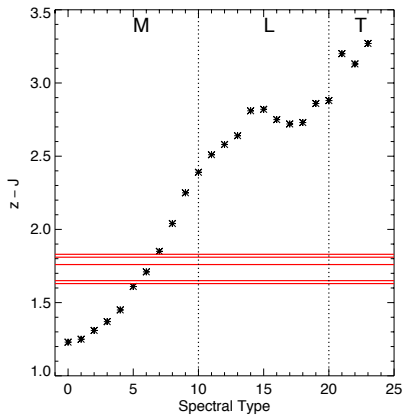}
  \caption{$z-J$ colours of the five T--type candidates as a function
of spectral type (red solid lines). The typical colours of M, L, and
T dwarfs from Sloan \citep{hawley02,west08,schmidt10b} are marked as
asterisks. The M, L, and T regions are delineated by vertical dotted
lines.}
  \label{fig_dT_PM:SpT_zmj}
\end{figure}

To break this degeneracy, we determined their $z-J$ colours and find 
these to be in the range 1.63--1.83$\pm$0.12 mag 
(Fig.\ \ref{fig_dT_PM:SpT_zmj}), typical of M4--M7 dwarfs 
\citep{hawley02,west05} but inconsistent with the expected much 
redder colours of L and T dwarfs 
\citep[$>$2.5 mag;][]{pinfield08,zhang09,schmidt10b}. 
Hence, on the basis of our new astrometric and photometric measurements 
we can rule out that these five candidates are young, 5-Myr old, 
planetary mass objects belonging to USco.

\section{Discussion and outlook}
\label{USco_dT_PM:discussion}

Combining our new $z$ and $H$ photometry with our proper motion 
measurements, we conclude that the five candidates proposed by 
\citet{lodieu11c} as young, T--type candidates are not cool brown 
dwarf members of the USco association. Hence, up to now, no T--type brown 
dwarf has been confirmed astrometrically and spectroscopically in this 
region. Overall, there are no young T-types confirmed spectroscopically 
in young star-forming regions, except the object reported by 
\citet{marsh10} but questionned by \citet{alves10}. No astrometric
confirmation is available and it will remain hard in $\rho$ Ophiuchus 
due to the small mean motion of members of this region.

We note the low success rate of wide-field surveys of young populations 
involving methane filters. For example, all of our five candidates are 
rejected after obtaining second epoch imaging and additional photometry. 
\citet{burgess09} reported three T--type candidates but later rejected two
of them using optical imaging. Similarly, \citet{spezzi12b} reported
four T--type candidates from a methane imaging combined with $JHK$ 
photometry and rejected two of them (the third may be a non
member too) from
their positions in various colour-colour and colour-magnitude diagrams.
We believe it is essential that additional photometry (e.g.\ deep optical 
$i$ and $z$), spectroscopy or astrometry is obtained for the 22 candidate 
T-type members of $\rho$ Oph identified by \citet{haisch10} so that 
their nature can be more rigorously examined.

So, after eliminating the five T-type candidates here, we have only one 
photometric candidate left (UGCS J16064645$-$2231238) in our one square 
degree survey in USco \citep{lodieu11c}. This candidate is the only object 
found in our survey with $J$ fainter than $\sim$19 mag. We reach a 
5$\sigma$ limit of 21 mag, similar to the deep VISTA survey of 
0.78 square degrees in $\sigma$ Orionis of \citet{penya12a} where 
three T--type candidates were identified, although proper motions 
indicate that two of them are likely to be non-members. Hence, our 
results are consistent with the main conclusions of \citet{penya12a} 
that we may see a turnover of the mass function below 10--4 Jupiter 
masses \citep[depending on the isochrones (NextGen, DUSTY, BT-Settl)
used and the age elected for USco (5 or 10 Myr)][]{baraffe98,chabrier00c,allard12} 
unless young T--type brown dwarfs are fainter than predicted by 
state-of-the-art models.

The next step in our quest for the bottom of the stellar/substellar 
initial mass function in USco is to obtain deeper and wider imaging 
using a combination of $Z,Y,J$ passbands where the USco sequence can 
be clearly de-lineated from the general field population 
\citep{lodieu07a}. We have targeted over 10 square degrees in USco with 
the largest infrared camera in the world, VIRCAM \citep{dalton06}, 
installed on VISTA to address this fundamental question regarding the 
fragmentation limit \citep{low76,rees76}. Our results will be presented 
in a forthcoming paper.

\section*{Acknowledgments}
NL was funded by the Ram\'on y Cajal fellowship number 08-303-01-02
and the national program AYA2010-19136 funded by the Spanish ministry 
of Economy and Competitiveness (MINECO). We thank Nigel Hambly for his 
advice on proper motion measurement.

This work is based on observations made with the ESO New Technology 
telescope at the La Silla Paranal Observatory under programme 
ID 089.C-0854(A) in visitor mode, and with the Gran Telescopio Canarias 
(GTC), operated on the island of La Palma in the Spanish Observatorio del 
Roque de los Muchachos of the Instituto de Astrof\'isica de Canarias. 

\bibliographystyle{mn2e}
\bibliography{../../../AA/mnemonic,../../../AA/biblio_old}
\end{document}